# X-ray microtomographic visualization of *Escherichia coli* by metalloprotein overexpression

**Ryuta Mizutani,[a*] Keisuke Taguchi,[a] Masato Ohtsuka,[b] Minoru Kimura,[b] Akihisa Takeuchi,[c] Kentaro Uesugi[c] and Yoshio Suzuki[c]**

[a]*Graduate School of Engineering, Tokai University, Kitakaname, Hiratsuka, Kanagawa 259-1292, Japan,* [b]*School of Medicine, Tokai University, Bohseidai, Isehara, Kanagawa 259-1193, Japan, and* [c]*Research and Utilization Division, JASRI/SPring-8, Kouto, Sayo, Hyogo 679-5198, Japan. E-mail: ryuta@tokai-u.jp*

**Synopsis** Three-dimensional structures of *Escherichia coli* overexpressing a metalloprotein ferritin were visualized by synchrotron radiation microtomography.

**Abstract** This paper reports X-ray microtomographic visualization of the microorganism *Escherichia coli* overexpressing a metalloprotein ferritin. The three-dimensional distribution of linear absorption coefficients determined using a synchrotron radiation microtomograph with a simple projection geometry revealed that the X-ray absorption was homogeneously distributed, suggesting that every *E. coli* cell was labeled with the ferritin. The ferritin-expressing *E. coli* exhibited linear absorption coefficients comparable to those of phosphotungstic-acid stained cells. The submicrometer structure of the ferritin-expressing *E. coli* cells was visualized by Zernike phase contrast using an imaging microtomograph equipped with a Fresnel zone plate. The obtained images revealed curved columnar or bunching oval structures corresponding to the *E. coli* cells. These results indicate that the metalloprotein overexpression facilitates X-ray visualization of three-dimensional cellular structures of biological objects.

## 1. Introduction

Soft tissues of biological objects are composed of light elements, which produce little contrast in a hard X-ray image. It is difficult to visualize a structure of biological interest even when applying contrast-enhancing methods such as phase contrast techniques or X-ray fluorescence detection. This is because the inherent contrast of soft tissues observed with X-ray absorption, interference, or fluorescence does not necessarily highlight the structure of biological interest. In conventional microscopy, biological objects are labeled with dyes or fluorescent probes to distinguish target structures from other tissue constituents. Images of the target structure can be obtained by using photons or electrons that interact with the probe.

Likewise X-ray visualization of biological objects should be performed by labeling biological constituents with probes that effectively interact with X-rays, i.e., high atomic-number (high-Z) elements (Mizutani & Suzuki, 2012).

A number of methods for staining biological constituents with high-Z probes have been reported (Ananda *et al.*, 2006; Mizutani *et al.*, 2007; de Crespigny *et al.*, 2008; Metscher, 2009). The high-Z probe staining is primarily performed by immersing biological objects in the probe reagent solution so as to allow the high-Z element to adsorb to, bind to, or be deposited on the target structure. Therefore, the effectiveness of the staining depends on the probe reagent's ability to penetrate into the biological tissue. Since most staining procedures are performed under aqueous conditions, hydrophobic tissue components such as fats and lipid bilayers act as barriers to probe permeation and thus hinder the staining processes.

Although the high-Z probe staining has been performed by introducing the probe from outside, X-ray labeling can also be performed by genetically engineering biological organisms to express a reporter protein that interacts with X-rays. This paper reports X-ray microtomographic visualization of the microorganism *Escherichia coli* that overexpresses a metalloprotein ferritin in the bacterial cells. Ferritin, which is responsible for iron storage and release (Chiancone *et al.*, 2004), can be used as a reporter protein for radiographic analyses. Three-dimensional structures of *E. coli* were visualized with an imaging microtomograph equipped with a Fresnel zone plate. The obtained images indicated that the metalloprotein overexpression facilitates X-ray visualization of the three-dimensional structures of biological objects.

## 2. Materials and Methods

### 2.1. Expression of recombinant ferritin

A ferritin gene segment was amplified from a pYX-Asc plasmid carrying murine ferritin mFTH1 gene (Open Biosystems) by polymerase chain reaction (PCR) using KOD Plus polymerase (Toyobo). Oligonucleotide primers (5'-GGAATTCCATATGTACCCATACGATGTTCCAGATTACGCTATGACCACCGCGTCTCCCTC-3') and (5'- TACCAAGCTTTTAGCTCTCATCACCGTGTCC-3') were used for introducing an *Nde*I site and an HA-tag immediately before the start codon of the mFTH1 gene and an *Hin*dIII site after the stop codon. The gel-purified *Nde*I-*Hin*dIII fragment of the PCR product was ligated into the *Nde*I-*Hin*dIII gap of pET-17b vector (Novagen) so as to construct a plasmid pET-17b-HA-mFTH1. *E. coli* strain BL21(DE3) carrying pLysS (Novagen) was transformed with this expression plasmid.

The obtained *E. coli* recombinant was cultured at 37°C to a logarithmic phase in an LB medium supplemented with 50 μg/ml of ampicillin. The expression of the HA-mFTH1 ferritin protein was induced by adding 0.1 mM isopropyl-1-thio-β-D-galactopyranoside and further incubating at 16°C overnight. Cells were harvested by centrifugation and washed with 50 mM Tris-HCl (pH 8.0). The obtained cells were suspended in lysis buffer (50 mM Tris-HCl, 1.25 mM EDTA, 20 mM $MgCl_2$, pH 8.0) containing 1 mg/ml of lysozyme (Sigma-Aldrich) and 70 unit/ml of DNaseI (Takara Bio). The cells were lysed by three cycles of freeze-thaw treatments.

The lysates were centrifuged at 16,000 × $g$ for 5 min. The resultant supernatants were incubated at 25°C for 5 min or 100°C for 3 min to denature proteins and then separated by SDS-PAGE. Electrophoretic transfer to a nitrocellulose membrane (NitroPure, Micron Separations) was performed using a semi-dry transfer cell (Bio-Rad). The membrane was blocked in 3% (w/v) BSA dissolved in phosphate-buffered saline (10 mM sodium phosphate, 0.15 M NaCl, pH 7.4) and subsequently incubated with anti-HA tag antibody (GeneTex) followed by horseradish-peroxidase-conjugated goat anti-rabbit IgG (KPL). Immunostained proteins were detected using diaminobenzidine (Sigma-Aldrich).

**2.2. Electron probe micro-analysis**

An LB medium supplemented with 1 mM ferric ammonium sulfate was centrifuged to remove precipitates. The medium supernatant was used to induce the HA-mFTH1 ferritin. *E. coli* cells were harvested from the overnight culture at 16°C and washed with Tris-buffered saline (0.15 M NaCl, 20 mM Tris-HCl, pH 7.5). The obtained cells were suspended in a Tris-buffered saline containing 2%(w/v) glutaraldehyde and incubated at 4°C overnight for glutaraldehyde fixation. The obtained cell pellets were dehydrated in 70% ethanol and then in absolute ethanol.

The cells resuspended in ethanol were then dried under vacuum and coated with gold. Electron probe micro-analysis was performed using an EPMA1610 analyzer (Shimadzu) operating at 15 kV. Distributions of sulfur and iron were simultaneously determined under the following conditions: beam current of 30 nA, unit measurement duration of 100 ms/pixel, and pixel size of 0.1 μm. A back-scattered electron image of *E. coli* cells was acquired using a 1-nA beam.

**2.3. X-ray microtomography**

The cell pellets dehydrated in ethanol were transferred into borosilicate glass capillaries (W. Müller Glas) with an outer diameter of 0.4–0.6 mm. After the pellet was dried, epoxy resin (Petropoxy 154, Burnham Petrographics) was introduced into the capillaries to embed *E. coli* cells in the resin. The capillaries were then incubated at 90°C for 16 h to cure the epoxy resin. A cell pellet of a mock recombinant of *E. coli* BL21(DE3) carrying pLysS and pET-17b vectors was also prepared with the same procedure. Staining with phosphotungstic acid was performed by incubating cells at 16°C overnight in an LB medium containing 10 mM phosphotungstic acid (Alfa Aesar).

The simple-projection microtomographic analysis was performed at the BL20XU beamline (Suzuki *et al*., 2004) of SPring-8. The sample capillary was mounted on the microtomograph by using a brass fitting specially designed for the glass capillary sample. Absorption contrast radiographs were recorded with a CMOS-based X-ray imaging detector (AA50 and ORCA-Flash2.8, Hamamatsu Photonics) using monochromatic radiation at 8 keV. The viewing field was rectangular with dimensions of 1920 pixels horizontally × 1440 pixels vertically. The data acquisition conditions are summarized in Table 1. Since the viewing field of this microtomograph is limited to 0.71 mm in height, datasets of the sample with an approximate height of 1.7 mm were taken in three batches. Each dataset was recorded by displacing the sample by 0.65 mm along the sample rotation axis.

Microtomographic analysis using high-resolution imaging optics was performed at the BL47XU beamline of SPring-8, as described previously (Takeuchi *et al*., 2011). A Fresnel zone plate with outermost zone width of 50 nm and diameter of 155 μm was used as an X-ray objective lens. Zernike phase-contrast radiographs produced by 8-keV X-rays were recorded (Takeuchi *et al*., 2009) using a CCD-based X-ray imaging detector (AA40P and C4880-41S, Hamamatsu Photonics). The viewing field was circular with a diameter of 1400 pixels. The data acquisition conditions are summarized in Table 1.

The obtained radiographs were subjected to convolution-back-projection calculation using the program RecView (Mizutani *et al*., 2011; available from http://www.el.u-tokai.ac.jp/ryuta/) accelerated with CUDA parallel-computing processors. Each microtomographic slice perpendicular to the sample rotation axis was reconstructed with this calculation. The spatial resolution of the reconstructed image (Table 1) was estimated by using three-dimensional square-wave patterns (Mizutani *et al*., 2010a).

## 3. Results

### 3.1. Expression of recombinant ferritin in *E. coli*

Supernatant of cell lysate of *E. coli* BL21(DE3) recombinant expressing the HA-mFTH1 ferritin protein was subjected to SDS-PAGE analysis (Fig. 1). Major bands observed at 24 kDa correspond to the molecular mass of the HA-mFTH1 polypeptide. Subsequent western-blot analysis using an anti-HA-tag antibody revealed that bands at 24 kDa as well as those remaining closer to the top of the gel contained HA-tagged proteins. The incubation at 100°C prior to SDS-PAGE raised the intensity of the 24-kDa band but slightly reduced the intensity of the band at the top of the gel. These results indicate that the HA-mFTH1 ferritin is resistant to detergent denaturation at room temperature. It has been reported that ferritins purified from chicken liver and horse spleen were observed as monomer bands after 100°C denaturation, but observed as high molecular-mass bands when the denaturation was performed at 24°C (Passaniti & Roth, 1989). The native form of ferritin is composed of 24 subunits (Chiancone *et al*., 2004). Therefore, it was suggested that the HA-mFTH1 ferritin was expressed in the same conformation as that of native ferritins. The total expression level of the HA-mFTH1 ferritin was estimated to be 30–50 mg protein per 1 mL wet cell volume.

The iron content of cell lysate of the ferritin recombinant that expressed 30 mg ferritin per 1 mL wet cell volume was determined (Fish, 1988) to be 13 mg iron per 1 mL wet cell volume. This corresponds to the incorporation of 4100 iron atoms into one 24-mer of the HA-mFTH1 ferritin, indicating that the HA-mFTH1 ferritin incorporated iron atoms, as has been reported for native ferritins (4500 irons per ferritin; Chiancone *et al*., 2004).

The recombinant cells were subjected to electron probe micro-analysis (Fig. 2). The outer shape of cells was determined from a back-scattered electron image, giving approximate dimensions of 0.4–0.6 μm in diameter and 2.0–3.5 μm in length. Some cells were observed as curved columns with a longer length. S $K\alpha$ and Fe $K\alpha$ emissions were mapped to determine overall protein and ferritin distributions. The mapping sample was flattened so as to allow electrons to penetrate into the sample. Although the Fe $K\alpha$ map indicated inhomogeneous distributions of iron, small spots that emitted strong X-rays should correspond to inorganic precipitates. Besides these spots, the Fe $K\alpha$ map showed a weak and widespread distribution of iron, which coincided with that of sulfur. This suggested that iron presumably derived from the medium was incorporated into the recombinant cells expressing the HA-mFTH1 ferritin.

### 3.2. Microtomographic analysis of *E. coli*

Structures of the ferritin-expressing recombinant and mock recombinant were visualized with synchrotron radiation microtomography. Figs. 3a-d show cross sections of *E. coli* pellets embedded in glass capillaries. The *E. coli* pellet of the ferritin-expressing recombinant prepared with the iron supplementation in the LB medium (Fig. 3a) exhibited a linear absorption coefficient of $10.6 \pm 1.4$ cm$^{-1}$ at 8 keV, while the surrounding resin exhibited $6.1 \pm 1.4$ cm$^{-1}$. The ferritin-expressing recombinant without the iron supplementation (Fig. 3b) exhibited an absorption coefficient of $9.4 \pm 1.6$ cm$^{-1}$, suggesting that the ferritin-expressing recombinant can accumulate iron from the intact LB medium. A pellet of the mock recombinant prepared with the iron supplementation (Fig. 3c) showed an absorption coefficient of $6.5 \pm 1.7$ cm$^{-1}$. The difference in X-ray absorption coefficients between the ferritin-expressing recombinant and mock recombinant both supplemented with iron is ascribable to the HA-mFTH1 gene incorporation. The absorption coefficient difference between the ferritin-expressing recombinant and mock recombinant is equivalent to an iron content of 14 mg/mL cell volume, which coincides with the iron content estimated above. Fig.

3e shows a histogram of the absorption coefficient of the *E. coli* pellet shown in Fig. 3a. A single peak in this histogram with a mean coefficient of 10.6 cm$^{-1}$ and a full width at half maximum of 3.2 cm$^{-1}$ can be approximated by a normal distribution with a mean iron content of 4300 atoms per ferritin and a standard deviation of 1400 irons per ferritin.

Microtomographic analysis was also performed for the mock recombinant pellet stained with phosphotungstic acid, which has been reported as a labeling reagent for biological tissues (Metscher, 2009). The absorption coefficient of the tungsten-stained *E. coli* was determined to be 8.4 ± 1.7 cm$^{-1}$ (Fig. 3d), which is comparable to the value obtained for the ferritin-expressing recombinant.

### 3.3. Three-dimensional structure of *E. coli* cells

Fig. 4a shows the three-dimensional structure of the ferritin-expressing recombinant pellet embedded in the glass capillary. The three-dimensional distribution of linear absorption coefficients revealed that the X-ray absorption was homogeneously distributed, suggesting that every *E. coli* cell was labeled with the HA-mFTH1 ferritin. The submicrometer structure of the ferritin-expressing recombinant cells was visualized by Zernike phase contrast using a microtomograph equipped with a Fresnel zone plate (Fig. 4b). The spatial resolution of the three-dimensional image was estimated to be between 150 nm and 200 nm using a microfabricated test object (Mizutani *et al*., 2010a). Although structural distortions can be introduced by the sample drying procedure (Mizutani & Suzuki, 2012), curved columns or bunching ovals with a cross-section diameter of 0.3–0.5 μm were observed in the three-dimensional image of the ferritin-expressing recombinant. This structure corresponds to the outer shape of *E. coli* observed in a backscattered electron image shown in Fig. 2a. Similar structures were also observed in the three-dimensional image of the mock recombinant stained with phosphotungstic acid (Fig. 4c). A stereo drawing of the phase contrast image of the ferritin-expressing recombinant is shown in Fig. 5. Some of the *E. coli* cells indicated subcellular localization of the X-ray phase shift. We suggest that these subcellular structures correspond to ferritin localization. These results indicate that the ferritin expression can be used for the visualization of three-dimensional cellular and subcellular structures of biological objects.

### 4. Discussion

Since hard X-rays hardly interact with light elements composing biological soft tissues, structures labeled with high-Z elements can be effectively visualized in radiographs without any clearing procedure. The high-Z element labeling has been achieved by introducing high-Z probes from outside, though the ability of probe reagents to permeate tissues has limited their application. The results obtained in this study indicate that the overexpression of ferritin gives sufficient contrast for X-ray visualization of biological objects. This method can be applied to any biological organisms that can be genetically engineered.

It has been reported that ferritin provides relaxivity contrast in magnetic resonance (MR) images (Genove *et al*., 2005; Deans *et al*., 2006; Cohen *et al*., 2007). Overexpression of the ferritin in transgenic mice allowed MR imaging of visceral structures (Cohen *et al*., 2007). However, MR imaging has not attained cellular resolution. By contrast, X-ray microtomography can visualize three-dimensional structures at subcellular resolution (Mizutani *et al*., 2010b). Overexpression of the ferritin under the control of a neuron specific promoter such as SYN1 promoter (Leypoldt *et al*., 2002) could facilitate visualization of neurons in X-ray images, like visualization of bones in clinical radiographs. This would allow microtomographic analysis of the three-dimensional network of neurons responsible for brain functions. Although the expression levels of probe proteins in transgenic animals would be much lower than that of the ferritin in *E. coli*, X-ray fluorescence microtomography has revealed three-dimensional iron distributions on the μg/g scale (Kim *et al*., 2006). We expect that the element-specific visualization methods, such as iron-edge difference microtomography and fluorescence microtomography, will be used for identifying iron labeled organisms in future. The results obtained in the present study suggest that ferritin overexpression combined with high-resolution X-ray microtomography can visualize three-dimensional structures of biological tissues at subcellular resolution.

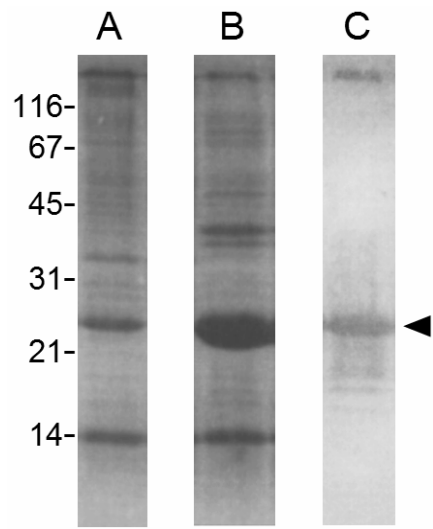

**Figure 1** SDS-PAGE (A, B) and western blot (C) analyses of supernatant of ferritin-expressing *E. coli* lysates. Lanes A and B were stained with Coomassie Brilliant Blue, and lane C with anti-HA-tag antibody. The HA-mFTH1 ferritin was observed primarily at 24 kDa, indicated with an arrow head, as well as at the top of the gel. Lanes A and C: supernatant denatured at 25°C prior to SDS-PAGE; lane B: supernatant denatured at 100°C. Molecular mass markers are indicated with labels.

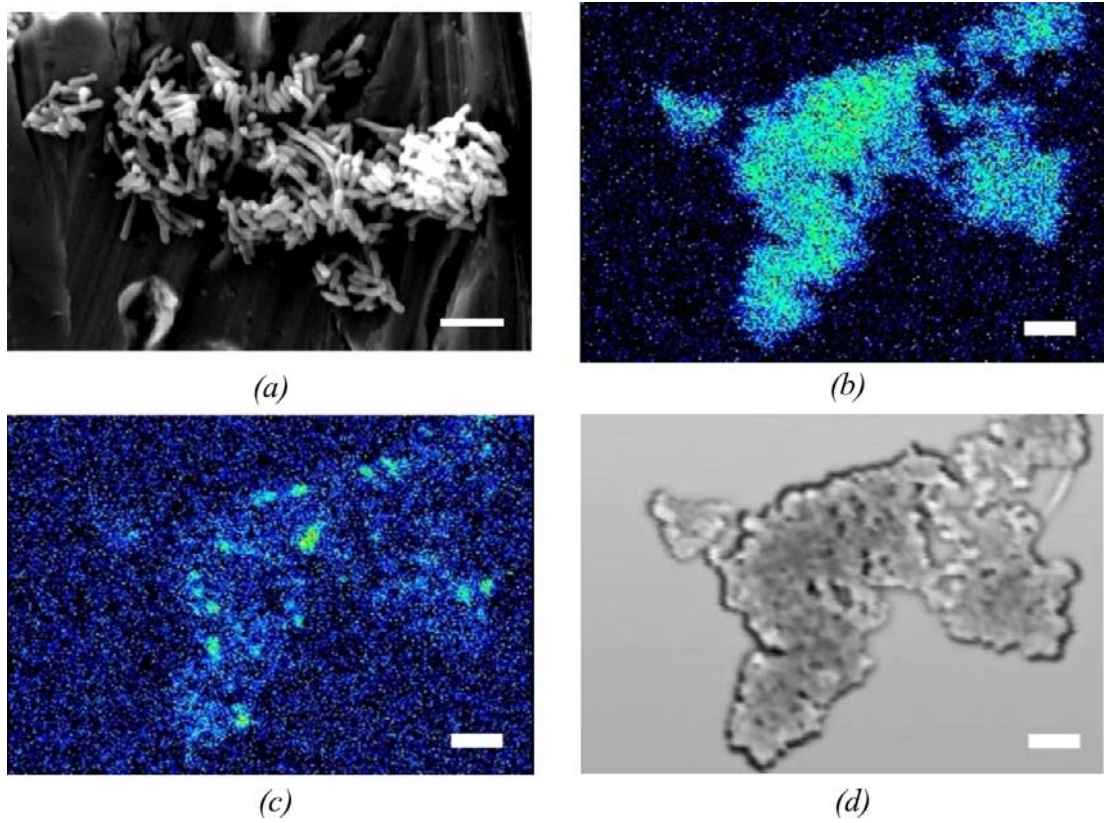

**Figure 2** Ferritin-expressing recombinant cells visualized with electron probe micro-analyzer. Scale bars: 5 μm. Back-scattered electron image (a) revealed outer shape of cells. A flattened sample was subjected to mapping of sulfur $K\alpha$ (b) and iron $K\alpha$ (c) distributions. A back-scattered electron image of the flattened sample is shown in panel (d).

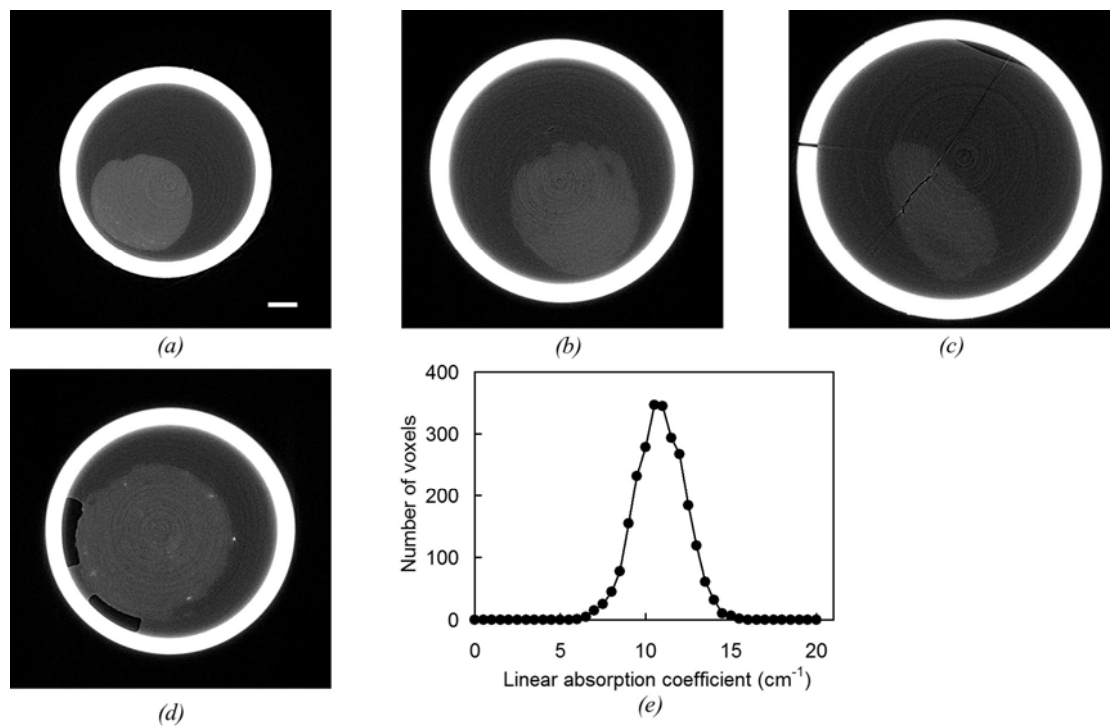

**Figure 3** Microtomographic cross sections of *E. coli* pellets embedded in glass capillaries. Linear absorption coefficients are shown in gray scale from 0 cm$^{-1}$ (black) to 30 cm$^{-1}$ (white). Scale bar: 50 μm. (a) Ferritin-expressing recombinant with iron supplementation in LB medium. (b) Ferritin-expressing recombinant without the iron supplementation. (c) Mock recombinant with the iron supplementation. (d) Mock recombinant stained with phosphotungstic acid. (e) Histogram of absorption coefficients of *E. coli* pellet shown in panel (a).

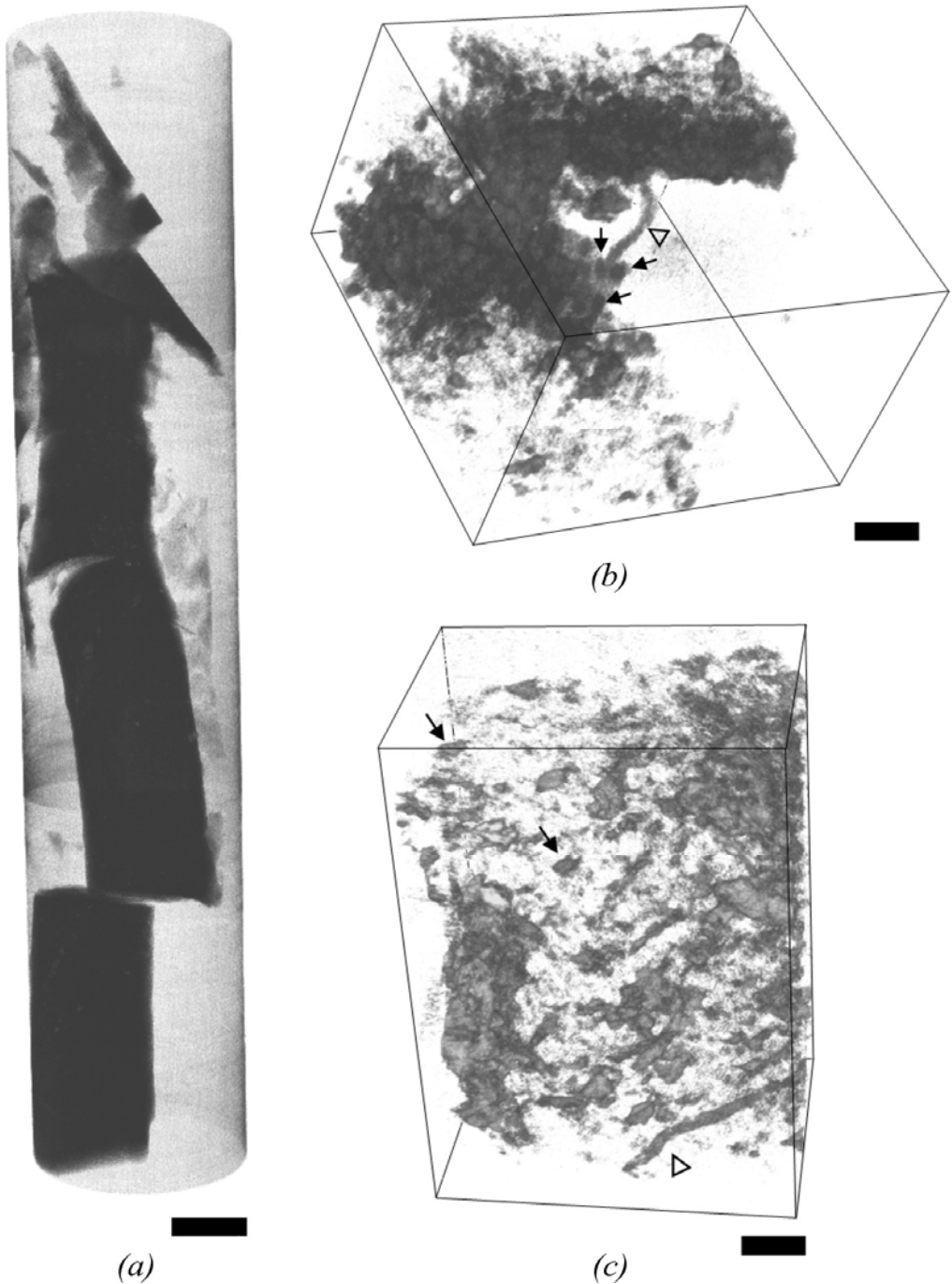

**Figure 4** (a) Three-dimensional visualization of a cell pellet of ferritin-expressing recombinant with simple projection microtomograph. Capillary voxels surrounding the cell pellet were removed. Linear absorption coefficients are rendered from 8 cm$^{-1}$ (white) to 50 cm$^{-1}$ (black). Scale bar: 100 μm. (b) Three-dimensional visualization of cellular structures of ferritin-expressing recombinant visualized by Zernike phase contrast using a microtomograph equipped with a Fresnel zone plate. The recombinant cells were observed as curved columns (triangle) and ovals (arrows). This image was taken near the upper end of the cell pellet shown in panel (a). Scale bar: 2 μm. (c) Structures of mock recombinant stained with phosphotungstic acid. Scale bar: 2 μm.

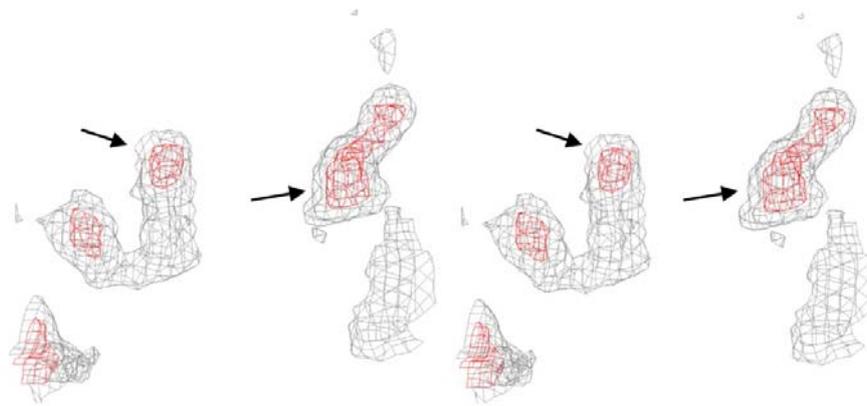

**Figure 5** Stereo drawing of three-dimensional phase shift distribution of ferritin-expressing recombinants located near the lower edge of Fig. 4b. X-ray phase shift was contoured at 2.5 (gray) and 3.2 (red) times the standard deviation of the entire three-dimensional image. Subcellular localizations are indicated with arrows. Grid size: 87.2 nm.

**Table 1** Conditions for microtomographic data collection

|  | Absorption contrast | Zernike phase contrast |
| --- | --- | --- |
| Beamline | BL20XU | BL47XU |
| X-ray energy (keV) | 8.0 | 8.0 |
| Pixel size (nm) | 492 × 492 [a] | 21.8 × 21.8 [a] |
| Viewing field (pixels) | Rectangular | Circular |
|  | 1920 × 1440 [a] | 1400 in diameter |
| Rotation/frame (degrees) | 0.10 | 0.05 |
| Exposure/frame (ms) | 150 | 500 |
| Frame/dataset | 1800 | 3600 |
| Dataset acquisition time (min) | 8 | 55 |
| Spatial resolution (nm) | 800-900 [b] | 150-200 |

[a] Width × height
[b] Mizutani *et al*. (2010c).

**Acknowledgements** We thank Yasuo Miyamoto and Kiyoshi Hiraga (Technical Service Coordination Office, Tokai University) for assistance with the electron probe micro-analyses and preparation of brass fittings for tomographic data collection. We thank the Department of Molecular Biology, Education and Research Support Center, Tokai University for assistance with DNA sequencing. This work was supported in part by a Grant-in-Aid for Scientific Research from the Japan Society for the Promotion of Science (no. 21611009). The synchrotron radiation experiments were performed at SPring-8 with the approval of the Japan Synchrotron Radiation Research Institute (JASRI) (proposal nos. 2011A0034, 2011B0034, 2011B0041, and 2012A0034).